\documentclass{openjournal}

\usepackage{lipsum}

\usepackage{xcolor}
\usepackage{textgreek}
\usepackage[utf8]{inputenc}
\usepackage[english]{babel}
\usepackage{needspace}

\usepackage{hyperref}
\hypersetup{
    unicode, 
    colorlinks=true,
    linkcolor=linkcolor,
    citecolor=linkcolor,
    filecolor=linkcolor,
    urlcolor=linkcolor,
}
\usepackage{color,colortbl}
\definecolor{linkcolor}{rgb}{0.0,0.3,0.5}
\usepackage{tensind}
\tensordelimiter{?}
\DeclareGraphicsExtensions{.bmp,.png,.jpg,.pdf}
\usepackage{verbatim}
\usepackage[normalem]{ulem}
\usepackage{orcidlink}
\usepackage{soul}

\def\gax{\mathrel{\raise.3ex\hbox{$>$}\mkern-14mu\lower0.6ex\hbox{$\sim$}}}
\def\lax{\mathrel{\raise.3ex\hbox{$<$}\mkern-14mu\lower0.6ex\hbox{$\sim$}}}
\def\gtorder{\mathrel{\raise.3ex\hbox{$>$}\mkern-14mu
             \lower0.6ex\hbox{$\sim$}}}
\def\ltorder{\mathrel{\raise.3ex\hbox{$<$}\mkern-14mu
             \lower0.6ex\hbox{$\sim$}}}
             
\graphicspath{ {./figs/} }

\begin{document}
\title{The Unhurried Universe: A Continued Search for Long Term Variability in ASAS-SN}

\author{Sydney Petz}
\email{petz.16@osu.edu}
\affiliation{Department of Astronomy, The Ohio State University, 140 W 18th Avenue, Columbus, OH 43210, USA}

\author{C. S. Kochanek} 
\affiliation{Department of Astronomy, The Ohio State University, 140 W 18th Avenue, Columbus, OH 43210, USA}
\affiliation{Center for Cosmology and AstroParticle Physics, 191 W Woodruff Avenue, Columbus, OH 43210, USA}

\author{K. Z. Stanek} 
\affiliation{Department of Astronomy, The Ohio State University, 140 W 18th Avenue, Columbus, OH 43210, USA}
\affiliation{Center for Cosmology and AstroParticle Physics, 191 W Woodruff Avenue, Columbus, OH 43210, USA}

\author{Benjamin J. Shappee}
\affiliation{Institute for Astronomy, University of Hawaii, 2680 Woodlawn Drive, Honolulu, HI 96822, USA}

\author{Subo Dong}
\affiliation{Department of Astronomy, School of Physics, Peking University, 5
Yiheyuan Road, Haidian District, Beijing 100871, People's Republic of China}
\affiliation{The Kavli Institute for Astronomy and Astrophysics, Peking University, 5 Yiheyuan Road, Haidian District, Beijing 100871, People's Republic of China}
\affiliation{National Astronomical Observatories, Chinese Academy of Science, Beijing 100101, People's Republic of China}

\author{J. L. Prieto}
\affiliation{Instituto de Estudios Astrofísicos, Facultad de Ingeniería y Ciencias, Universidad Diego Portales, Avenida Ejercito Libertador
441, Santiago, Chile}
\affiliation{Millennium Institute of Astrophysics MAS, Nuncio Monseñor Sotero Sanz 100, Off. 104, Providencia, Santiago, Chile}

\author{Todd A. Thompson} 
\affiliation{Department of Astronomy, Ohio State University, 140 West 18th Avenue, Columbus, OH 43210, USA}
\affiliation{Center for Cosmology and Astro-Particle Physics, Ohio State University, 191 West Woodruff Ave., Columbus, OH 43210, USA}
\affiliation{Department of Physics, Ohio State University, 191 West Woodruff Ave., Columbus, OH 43210, USA}

\begin{abstract}
We search a sample of 5,685,060 isolated sources in the All Sky Automated Survey for SuperNovae (ASAS-SN) with $14.5<g<15$ mag for slowly varying sources with brightness changes larger than $\sim 0.03$~mag/year over 10 years. We find 426 slowly-varying systems. 
Of these systems, 200 are identified as variables for the first time, 226 are previously classified as variables, 
and we find equal numbers of sources becoming brighter and fainter.
Previously classified systems were mostly identified as semi-regular variables (SR), slow irregular variables (L), or unknown (MISC or VAR), as long time scale variability does not fit into a standard class.
Much like \cite{Petz25}, the sources are scattered across the color magnitude diagram and can be placed into 5 groups that exhibit distinct behaviors. There are also six AGN.
There are 262 candidates ($\sim$~62 percent) that also show shorter time scale periodic variability, mostly with periods longer than 10 days.
The variability of 66 of these candidates may be related to dust.
Combining the new slow variable candidates with the candidates from \cite{Petz25}, we have found a total of 1208 slow variables.

\end{abstract}


\maketitle

\section{Introduction}
\label{sec:intro}
Despite the growing number of high cadence All-Sky Surveys, such as the All-Sky Automated Survey for SuperNovae (ASAS-SN; \citealt{ Shappee2014, Kochanek2017, Jayasinghe2018}), the Zwicky Transient Facility (ZTF; \citealt{Bellm2014}), the Asteroid Terrestrial-impact Last Alert System (ATLAS; \citealt{Heinze2018}) and the upcoming Legacy Survey of Space and Time at the Vera Rubin Observatory (LSST; \citealt{Hambleton2023}), systematic exploration of the slowly varying universe has only recently begun.
Initial studies of long term variability \textemdash objects with slow brightness trends over timescales of years \textemdash have primarily focused on magnetic activity cycles in solar type stars \citep{Baliunas1990,Weis1994}, however more recent investigations of slow variability have focused on individual interesting systems.


Stellar mergers are one rare example, with an estimated rate of approximately one per year in the Galaxy \citep{Kochanek2014}. They show slow variability changes both before and after the merger event as seen in V1309 Scorpii \citep{Tylenda2011} and M31-LRN-2015  \citep{Blagorodnova2020}.
More recently, slower, steady brightening sources have been investigated, such as the slow brightening of WNTR23bzdiq/WTP19aalzlk \citep{Karambelkar25}, which resembles stellar merger transients on longer time scales and is proposed to potentially mark the onset of common-envelope evolution.
Other unique systems, such as the cool giant Gaia17bpp \citep{Tzanidakis2023} which faded $\sim 4.5$ mag over $\sim$6.5 years, make up a class of dramatic dimming events that span timescales of several months or years, often proposed to be caused by eclipsing dusty disks (e.g., \citealt{Rowan2021,Smith2021, Torres2022}). With an increasing number of discoveries of these long timescale dimming events, there has been an increasing effort to characterize their behavior at a population level \citep{Raquel25}.
Although other previous work has investigated the long term behavior of larger groups of objects such as active galactic nuclei (AGN) \citep{Yang2016}, Type III classical T Tauri stars (cTTS) \citep{Grankin2007}, and quasars \citep{MacLeod2012}, the slowly varying universe still remains largely unexplored. 

\citealt{Petz25} (referred to as PK25) explored the slow variability of $\sim$ 9 million sources with $13<g<14.5$ mag in ASAS-SN, finding 782 candidates displaying slow variability correlated with their positions on a Gaia DR3 $M_{G}$ and $B_{P}-R_{P}$ color magnitude diagram.
To find more instances of this unique type of long term variability, we use ASAS-SN to search for slowly-varying sources with $14.5<g<15$ mag.
We describe the candidate selection and analysis methods in \S\ref{sec:methods} and the results of our search in \S\ref{sec:results}.
We discuss possible follow up studies and survey extensions in \S\ref{sec:concl}.

\section{Methods}
\label{sec:methods}
Following PK25, we downloaded the light curves of the 5,685,060 sources with 14.5$<$g$<$15 mag from ASAS-SN Sky Patrol V2.0 \citep{Hart23} and intercalibrated the cameras and the V and g band data using a damped random walk Gaussian process for interpolation with the camera/filter calibration shifts fit as additional linear parameters (see \citealt{Kozlowski2010}). 
To only focus on long timescale variability, we computed seasonal medians for each light curve and fit them using both a linear and a quadratic function of time.
For the quadratic fits we define a ``slope'' $\Delta g/\Delta t$ where $\Delta g$ is the maximum magnitude difference over the span of the light curve $\Delta t$. 
Following PK25, we make an empirical cut at an absolute linear slope of 0.03 mag/year and discarded sources slopes less than 0.03 mag/year and sources with a $\Delta g$ less than 0.3 mag.
This left us with 26,518 candidates.

\begin{figure}[!tb]
    \centering
    \includegraphics[width=0.7\textwidth]{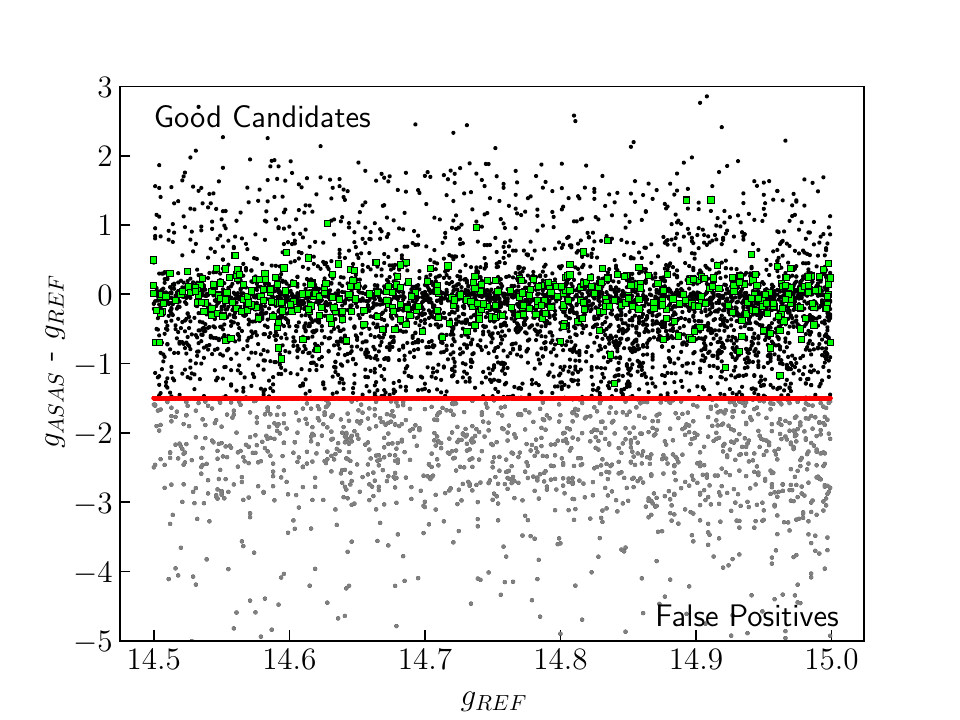}\\
    \caption{Distribution of candidates after the first bright star cut. We keep stars above the red solid line. Green squares show the candidates that make up our final sample.} 
    \label{fig:bscuts}
\end{figure}

Next we discard several classes of false positives.
The first class is a range of artifacts created by bright stars\textemdash in particular the bleed trails of very saturated stars can cause false positives for slow variability even when the target star is quite separated on the sky. 
Following PK25, we examine the current sample in the space and magnitude of nearby bright stars, and identify the same two bands of candidates, the lower band consisting of false positives close to nearby bright stars.
We keep candidates where the distance to the nearest bright star of magnitude $g$ is $R>0.2\exp(12.5-g)+70$ arcseconds, saturating with R=min(R,3600). This second empirical cut leaves us with 5819 candidates.
We also noticed that stars affected by this problem often show large differences between the median ASAS-SN magnitudes and their corresponding $g$ band REFCAT \citep{REFCAT} magnitudes (see Fig.~\ref{fig:bscuts}). 
To discard these sources, we only keep stars with median ASAS-SN magnitudes that are no more than 1.5 mag brighter than their REFCAT g magnitude.
This cut also eliminated the false positives near the celestial south pole found by PK25 that are caused by the field rotations created by ASAS-SN's equatorial mounts. This leaves us with 4563 candidates.
Next we removed high proper motion stars or stars near high proper motion stars. As in PK25, we eliminate stars with Gaia DR3 \citep{GaiaDR3} proper motions higher than 100 mas/year or where a nearby star with a flux ratio greater than 0.01 passes through the photometric aperture. This removes 14 candidates.
We visually examined the light curves and the corresponding ASAS-SN reference images of the remaining 4549, allowing us to remove any remaining light curves that have large magnitude uncertainties and are still being affected by nearby bright stars. 
Nearly all of the removed candidates have light curves with saturation issues from nearby bright stars where the median magnitudes of several observing seasons are artificially elevated.
This left us with the 426 candidates listed in Table~\ref{tab:cands}.
There are nearly equal numbers of sources becoming brighter (209) and fainter (217), so they are unlikely to be due to systematic drifts in the photometry.

We cross matched these sources with 
Gaia Alerts \citep{GaiaAlerts}, and the AAVSO VSX catalog \citep{AAVSO}, which contains variables identified by amateurs and surveys such as ASAS-SN, ATLAS, WISE \citep{Chen2018}, and ZTF.
We also cross match our final candidates with Gaia DR3 photometry \citep{GaiaDR3}, use distances from \cite{BJ2023}, and correct for extinction with \texttt{mwdust} \citep{Mwdust}, which is based on the dust maps of \cite{Drimmel}, \cite{Marshall2006}, and \cite{Green2019}. We identified likely active galactic nuclei (AGN) by matching the sources to the {\tt milliquas v8} catalog (\citealt{Flesch2023}) and used {\tt SIMBAD} (\citealt{Wenger2000}) to obtain general stellar classifications and spectral types if they were available.

To compare the optical and mid-IR variabilities, we extract Near-Earth Object Wide-field Infrared Survey Explorer (NEOWISE; \citealt{Mainzer2014}) light curves for each source using \cite{WISE-pull-lc}.
We combine points from the same day, and fit the W1 light curve and the W1$-$W2 color evolution with linear and quadratic functions of time.
Additionally, we use the Gaia DR3 light curves \citep{GaiaDR3}, to explore the changes in the G, $B_{p}$ and $R_{p}$ magnitudes.
We searched for periodicity on shorter time scales by first subtracting  the linear or quadratic trend from our initial fits and then use a Lomb-Scargle periodogram \citep{Lomb1976,Scargle1982} to search for their periods.
We do not report periods longer than the average observing season.
We keep the most significant period with a standard GLS false alarm probability $<$ 0.1 and examine the phased light curve for each candidate to remove spurious periods.






\section{Results}
\label{sec:results}
\begin{figure}[!tb]
    \centering
    \includegraphics[width=\textwidth]{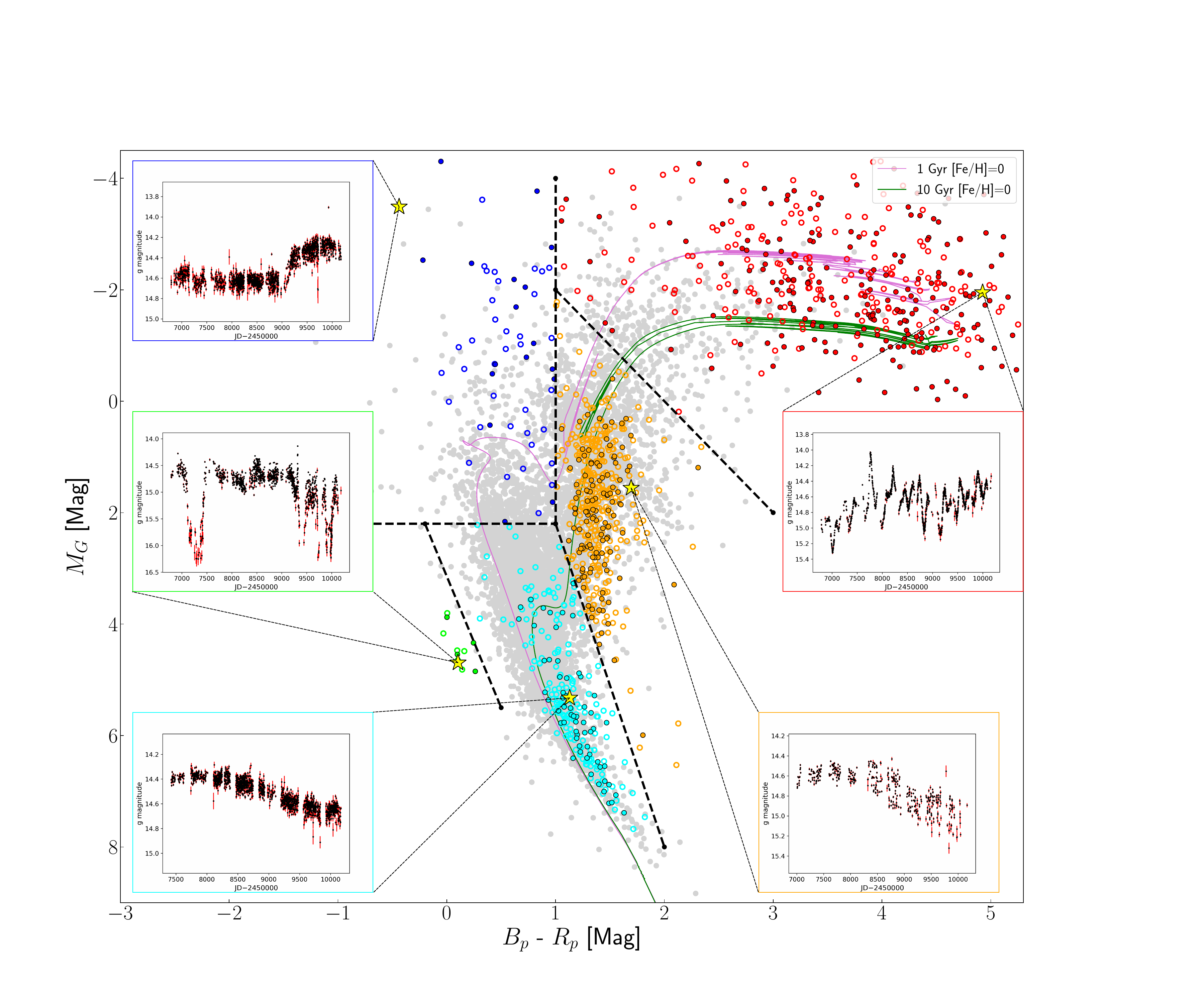}\\
    \caption{The Gaia DR3 $M_{G}$ and $B_{P} - R_{P}$ color-magnitude diagram of the final candidates divided into groups based on the dashed lines. The curves are Solar metallicity 1 and 10 Gyr MIST isochrones. The empty points show the CMD positions of candidates from PK25 and the filled in points show the new candidates from this work. One example of light curve is shown for each group.}
    \label{fig:cmdgroups}
\end{figure}

After searching through a sample of 5,685,060 objects within $14.5<g<15$ magnitudes, we find a total of 426 candidates exhibiting long term variability.
This includes 226 objects that were previously flagged as variables, and 200 that were not.
These candidates are displayed in Fig.~\ref{fig:cmdgroups} on an extinction corrected Gaia DR3 $M_{G}$ and $B_{P}-R_{P}$ color magnitude diagram (CMD) with Solar metallicity 1 Gyr and 10 Gyr MIST \citep{Paxton2018} isochrones to track stellar evolutionary stages.
As in PK25, we separate the candidates into five groups based on their position on the CMD, shown in Fig.~\ref{fig:cmdgroups}. The 5 groups are main sequence, subgiant/giants, AGB stars, luminous blue stars, and novae, with each group exhibiting different physical characteristics in their light curves.
Due to negative or missing distances in \cite{GaiaDR3, BJ2023}, 15 of the final candidates cannot be placed into a group.

\begin{figure}[!tb]
    \centering
    \includegraphics[width=\textwidth]{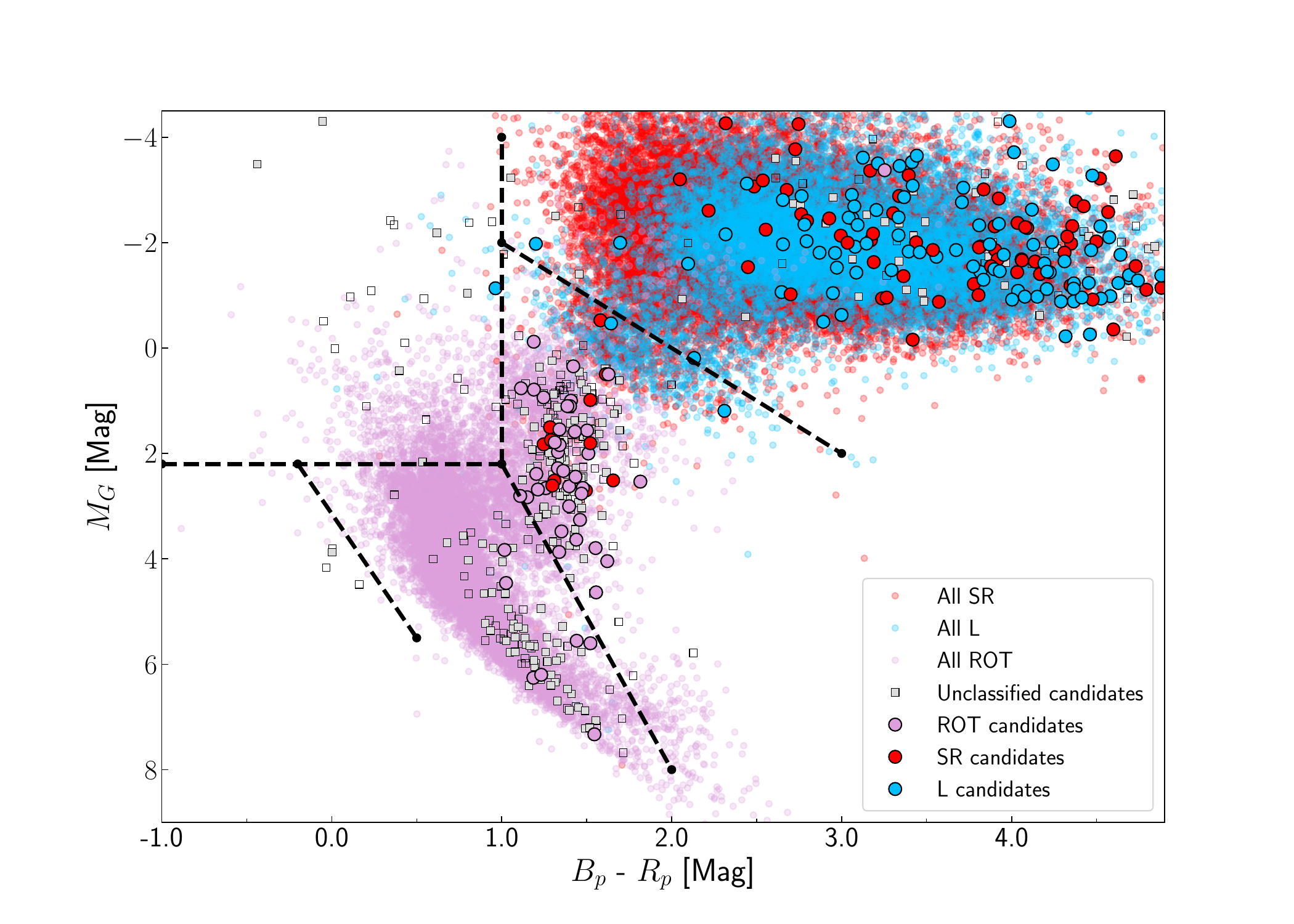}\\
    \caption{The Gaia DR3 $M_{G}$ and $B_{P} - R_{P}$ color-magnitude diagram of all variables in SkyPatrol V2.0 with $13<g<15$ mag that are classified as SR, L, MISC, and VAR (small points) as compared to the candidates with these classifications (large points). Previously unclassified candidates are marked by squares. The empty points show the CMD positions of candidates from PK25 and the filled in points show the new candidates from this work.}
    \label{fig:twins}
\end{figure}

The two most common types of known variables found in the sample are semi-regular (SR) and slow irregular (L) pulsating variables, with a total of 122 pulsating variables in all. Additionally, there are 59 classified as a variable of unknown type (MISC or VAR). This is not surprising, as this slow variability does not fit into a standard type.
We also match to 3 eclipsing variables, 14 rotating variables, 20 eruptive variables, and 6 cataclysmic variables. Only one candidate was classified as an eclipsing and pulsating variable, and 6 were identified as AGN in {\tt milliquas v8}. Fig.~\ref{fig:twins} compares the candidates previously classified as SR, L, and ROT (spotted star rotating variables)candidates to all
such variables in Sky Patrol v2.0 with $13 < g < 15$~mag in a CMD.
As seen with the brighter candidates in PK25, the previously classified SR and L variables in our sample are somewhat more luminous than the broader population. We see a much smaller number of ROT candidates in this fainter population.

\begin{figure}[!ht]
    \centering
    \includegraphics[width=0.7\textwidth]{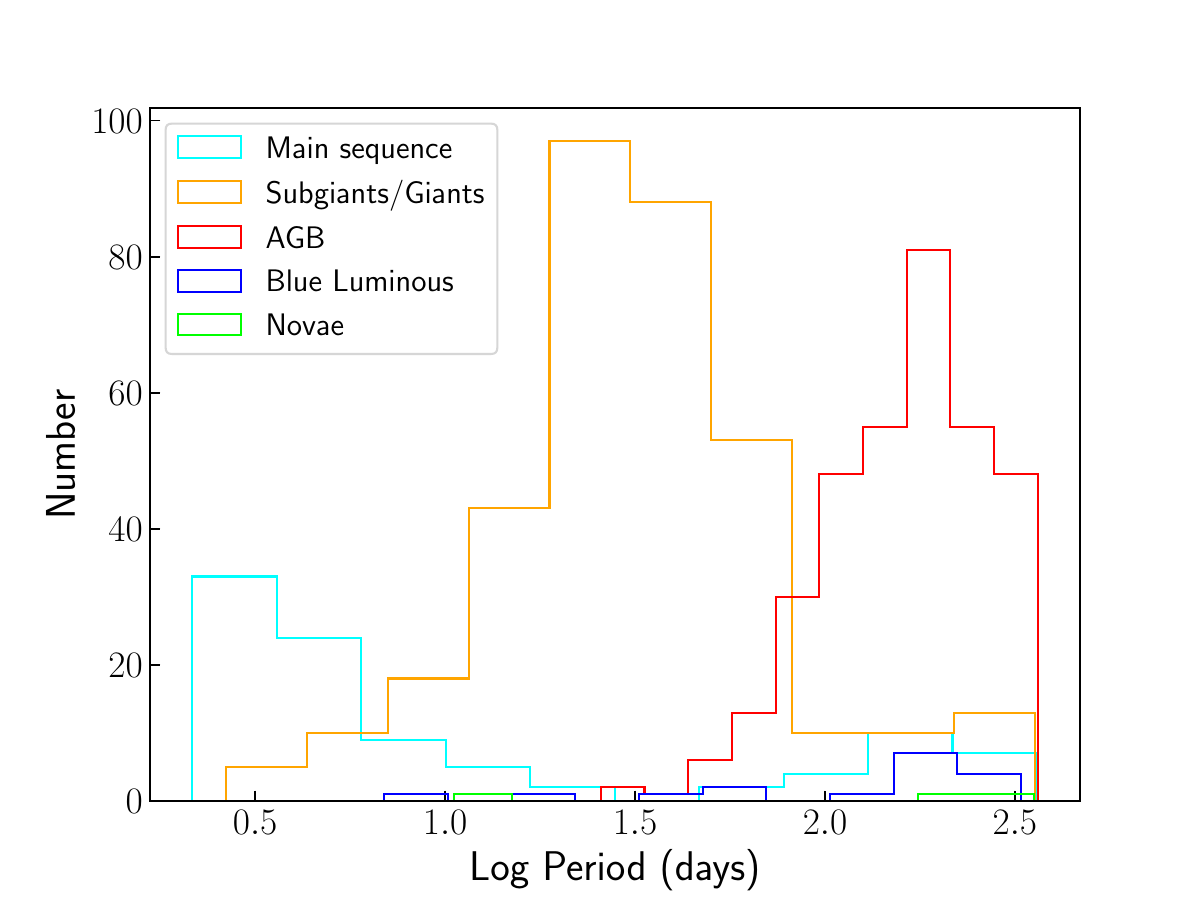}\\
    \caption{Total period distribution for the candidates by group with $13<g<15$ mag.}
    \label{fig:phist}
\end{figure}

Fig.~\ref{fig:phist} displays the period distribution for the periodic variables in each group including all candidates from both PK25 and this work.
Matching the trend set by the brighter candidates in PK25, the periods are essentially ordered by stellar size, with the main sequence group at short periods, the subgiant/giants at intermediate periods, and the AGB group at long periods. There are only a handful of periodic stars in the luminous blue and nova groups. Although, the overall distribution is still dominated by the subgiant/giant group, however there are a large number of AGB group candidates in this fainter sample.

\begin{figure}[!tb]
    \centering
    \includegraphics[width=0.5\textwidth]{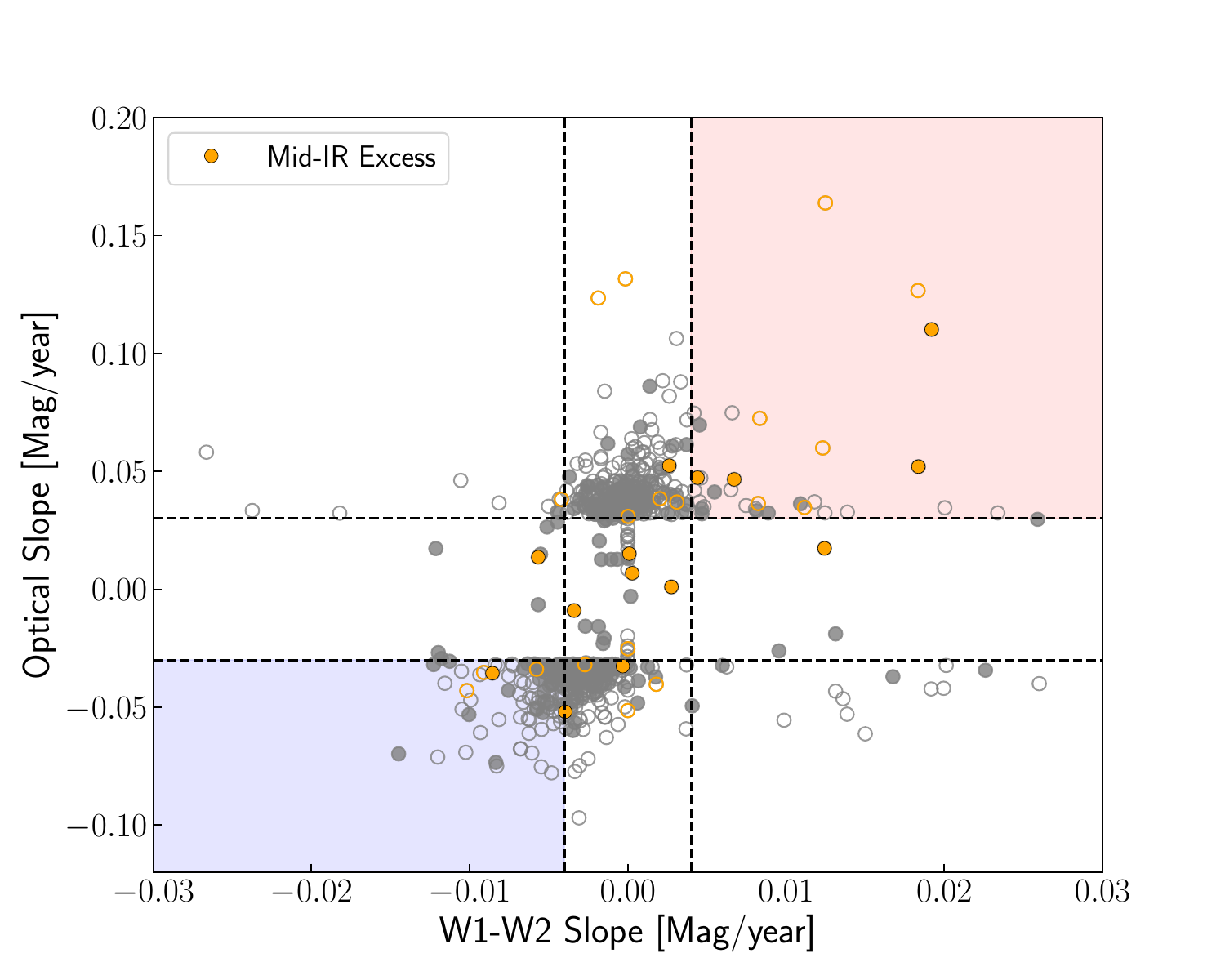}\includegraphics[width=0.5\textwidth]{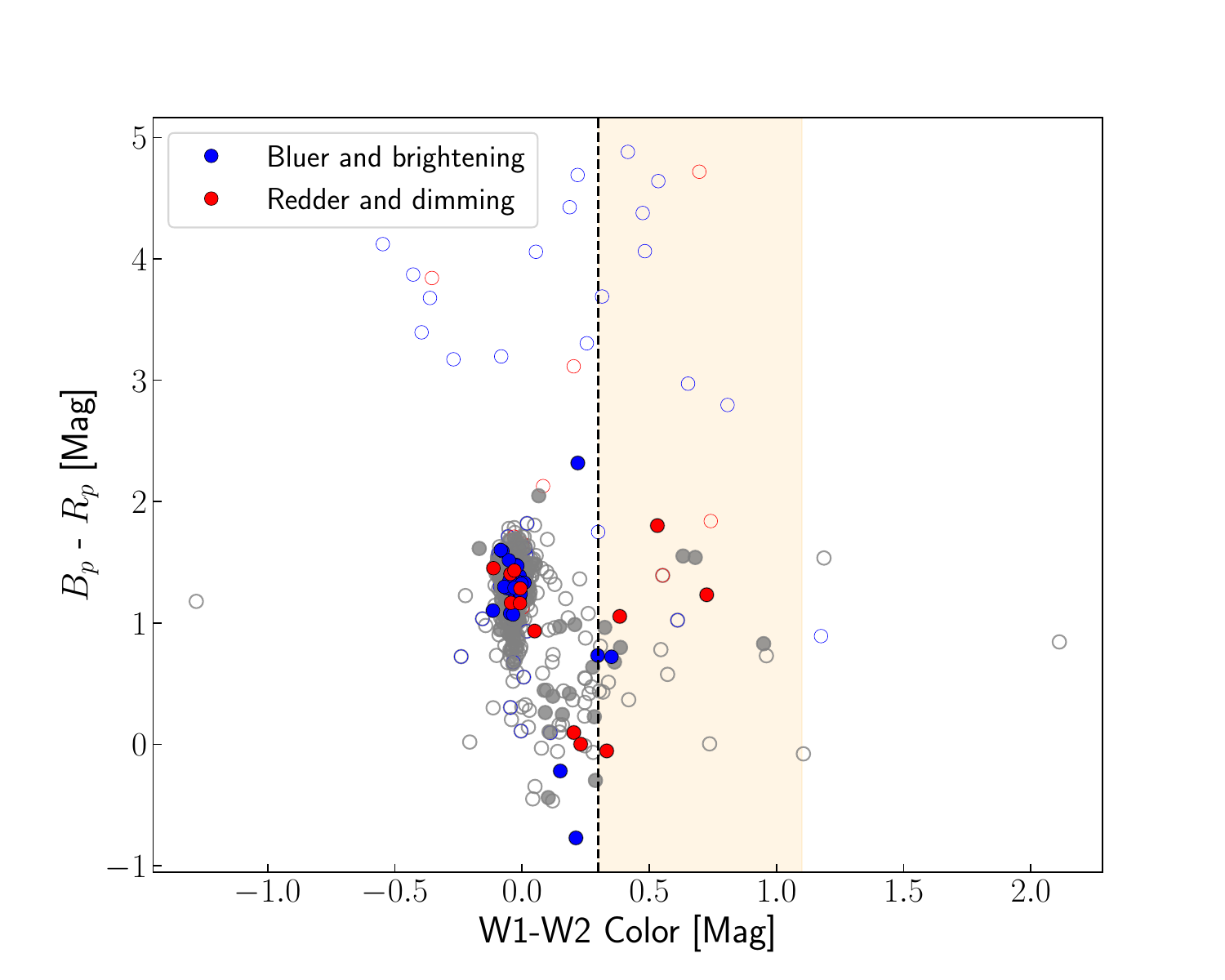}\\
    \caption{Left: Optical variability slope as a function of the W1$-$W2 color variability slope. Red and blue shaded regions indicates candidates that are likely getting redder and dimmer, or bluer and brighter. Right: $G_{BP} - G_{RP}$ color as a function of mean W1$-$W2 color. Candidates in the orange shaded region have a mid-IR excess due to dust emission. Candidates from the shaded region of each panel are shown with colored symbols on the other panel. The empty points show the candidates from PK25 and the filled in points show the new candidates from this work.}
    \label{fig:wise1}
\end{figure}

\begin{figure}[!tb]
    \centering
    \includegraphics[width=\textwidth]{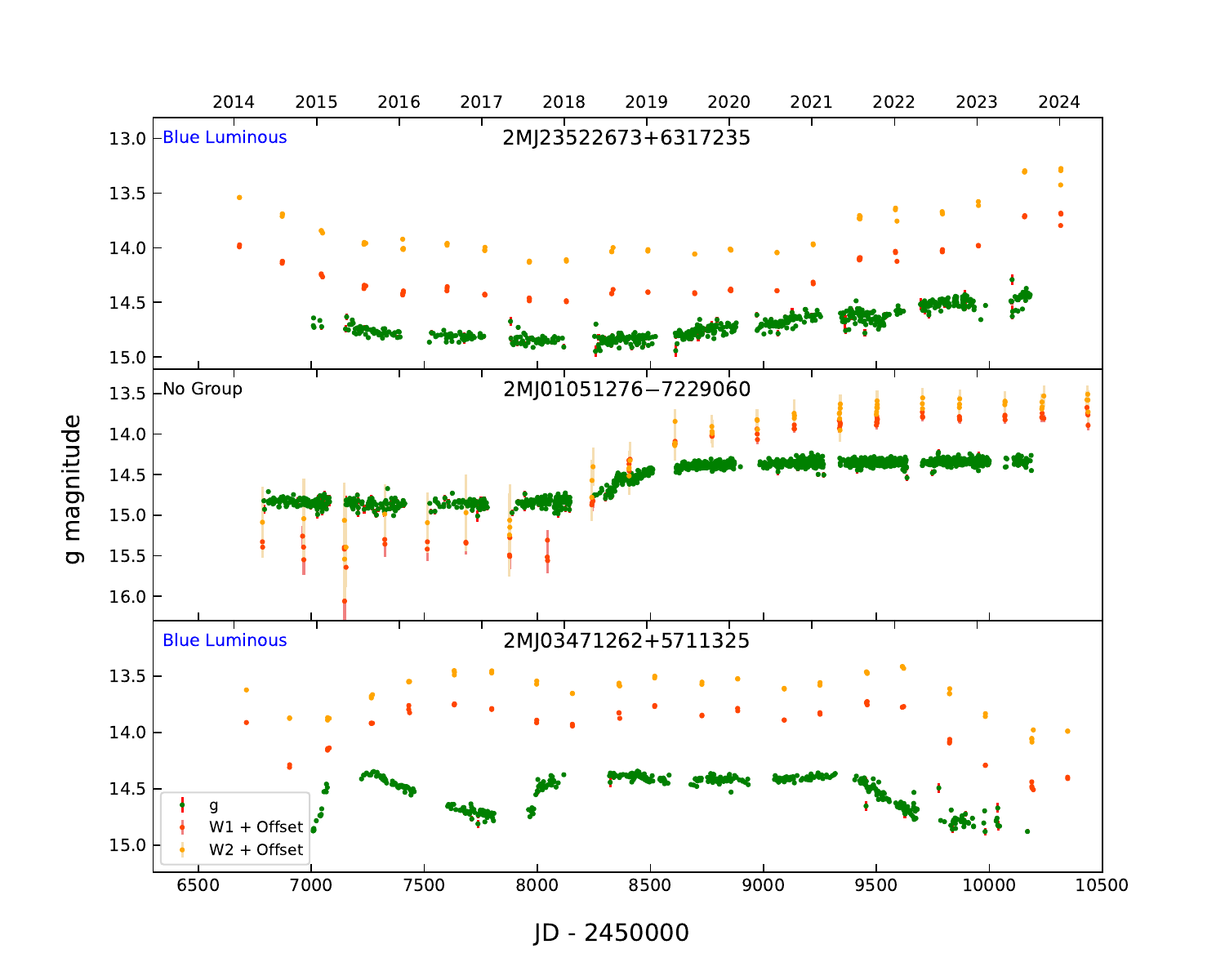}\\
    \caption{Comparisons of ASAS-SN and WISE light curves of candidates labeled by their 2MASS ID and CMD group.}
    \label{fig:wise2}
\end{figure}

To assess whether the optical variability of our candidates may be associated with dust, we examine their WISE light curves. Fig.~\ref{fig:wise1} shows two comparisons of the optical and WISE mid-IR properties of the candidates.  The right panel compares the optical extinction corrected mean $B_P-R_P$ color to the mean WISE $[W1]-[W2]$ color. Following PK25, we exclude stars with $[W1]<9$ mag where saturation starts to affect the colors. We flag stars redder than $[W1]-[W2] > 0.3$~mag as good candidates for stars with circumstellar dust because they have a significant mid-IR excess.  
The left panel compares the optical g-band light curve slope $s_g$ mag/year and the mid-IR $[W1]-[W2]$ color slope $s_{col}$ mag/year.  
Good candidates for variability driven by new dust formation should have $s_g>0.03$  mag/year (fading) and $s_{col}>0.004$ mag/year (becoming
redder), while variability driven by dust being destroyed or becoming more distant from the star should have the opposite trend. We identify 49 fading/brightening systems and 17 systems with a mid-IR excess as possible dust driven variability candidates, as shown in both panels in Fig.~\ref{fig:wise1}.
Fig.~\ref{fig:wise2} shows examples of (1) an unclassified blue luminous star with a mid-IR excess, (2) a star with a large brightening event in both the optical and infrared, and (3) Mis~V1325, a rapid irregular variable which shows a dust driven event at JD$-$ 2450000 $\sim$ 7250 where the optical fades and the mid-IR brightens.

\begin{figure}[!tb]
    \centering
    \includegraphics[width=0.5\textwidth]{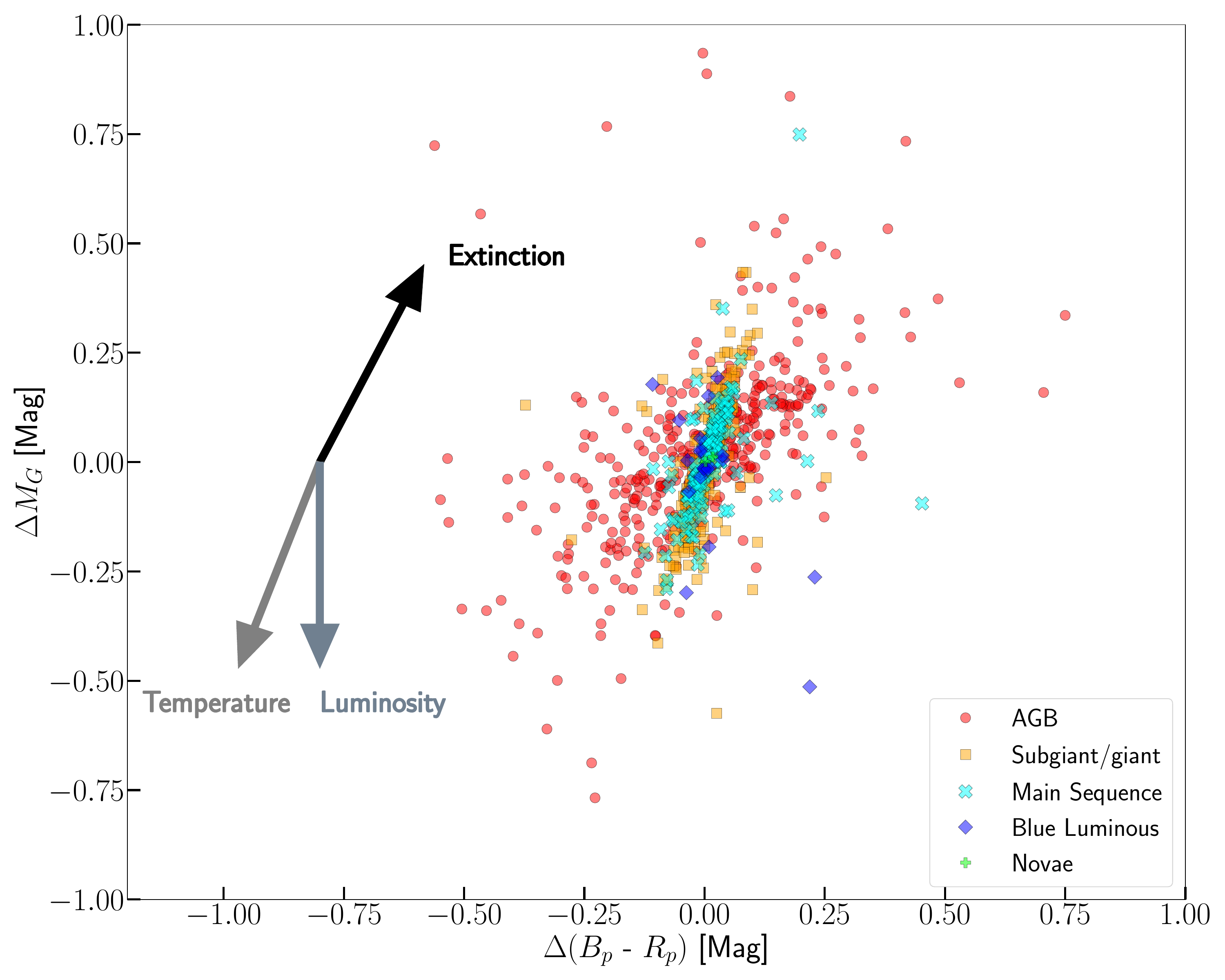}\hspace{-0.5em}%
    \includegraphics[width=0.5\textwidth]{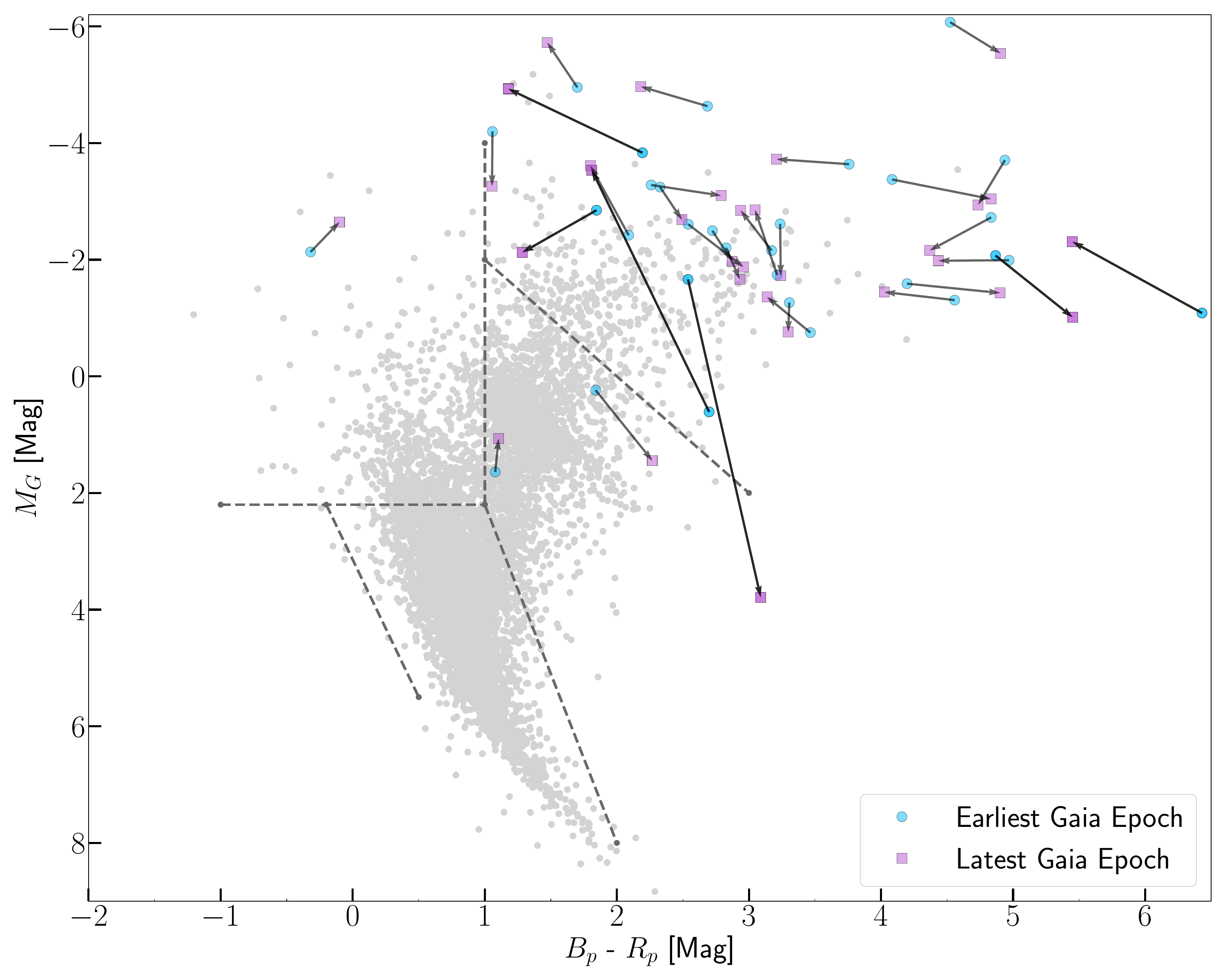}

    \caption{Left: $\Delta M_{G}$ and $\Delta(B_{p} - R_{p})$ of all candidates with Gaia DR3 epoch photometry. The solid arrows display the  directions in extinction ($\textrm{R}_V=3.1$), luminosity, and temperature. Candidates with an absolute $|\Delta M_{G}| > 1$ mag are excluded.
    Right: A Gaia CMD showing all candidates with $|\Delta M_{G}|$ or $|\Delta(B_{p} - R_{p})| > 0.5$ mag. Blue circle points are the earliest epoch of observation and purple square points are the latest epoch of observation.}
    \label{fig:gaia}
\end{figure}

\begin{figure}[!tb]
    \centering
    \includegraphics[width=0.5\textwidth]{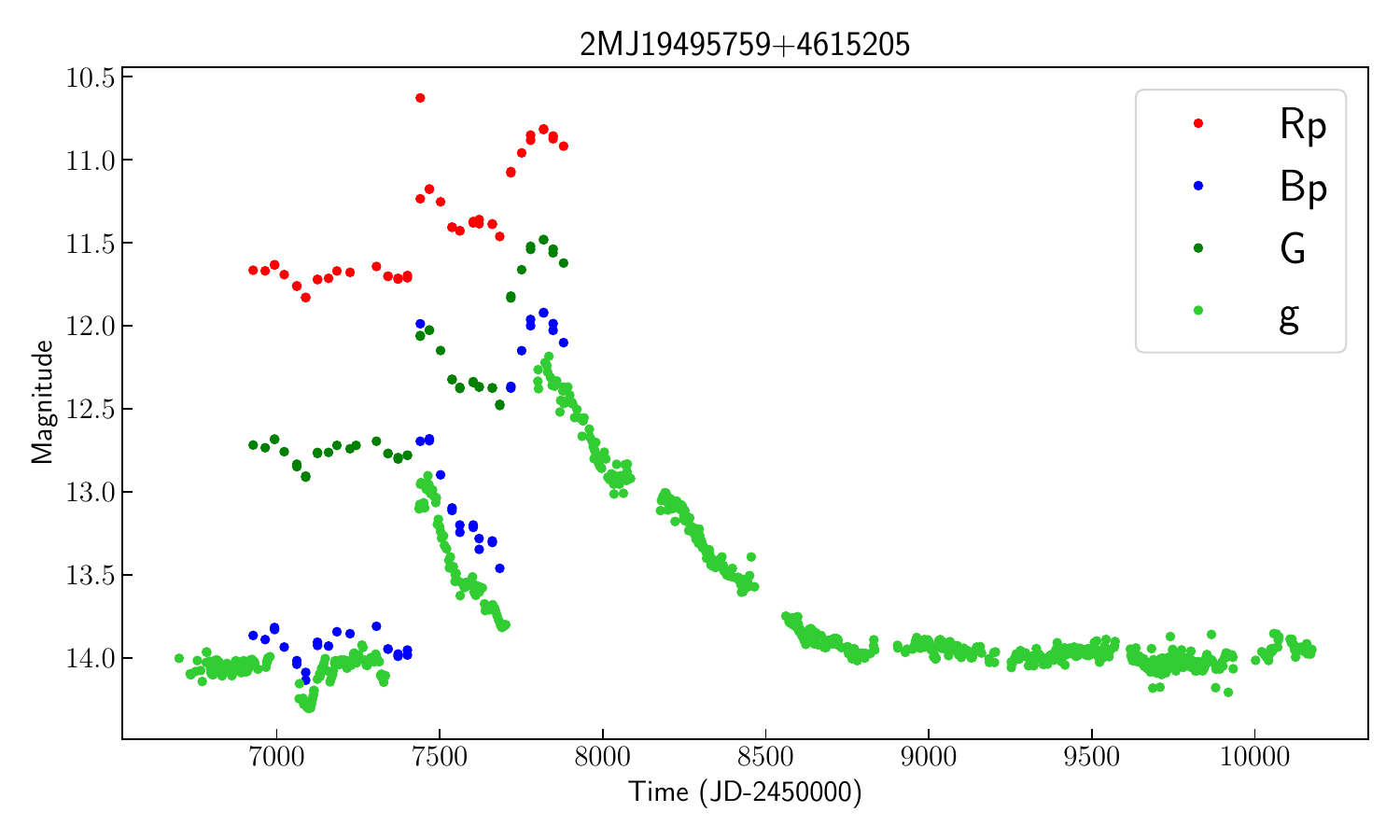}\hspace{-0.5em}
    \includegraphics[width=0.5\textwidth]{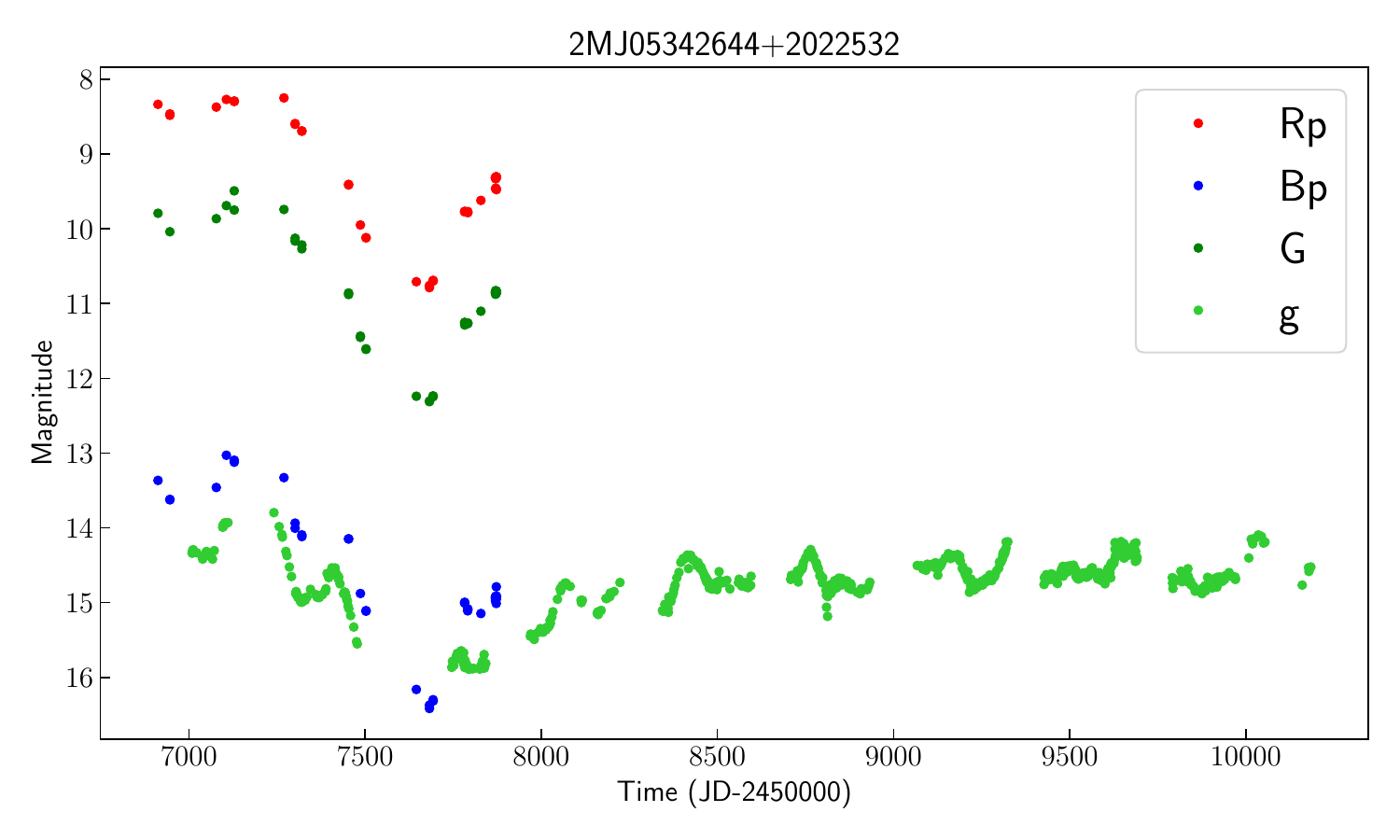}\\
    \includegraphics[width=0.5\textwidth]{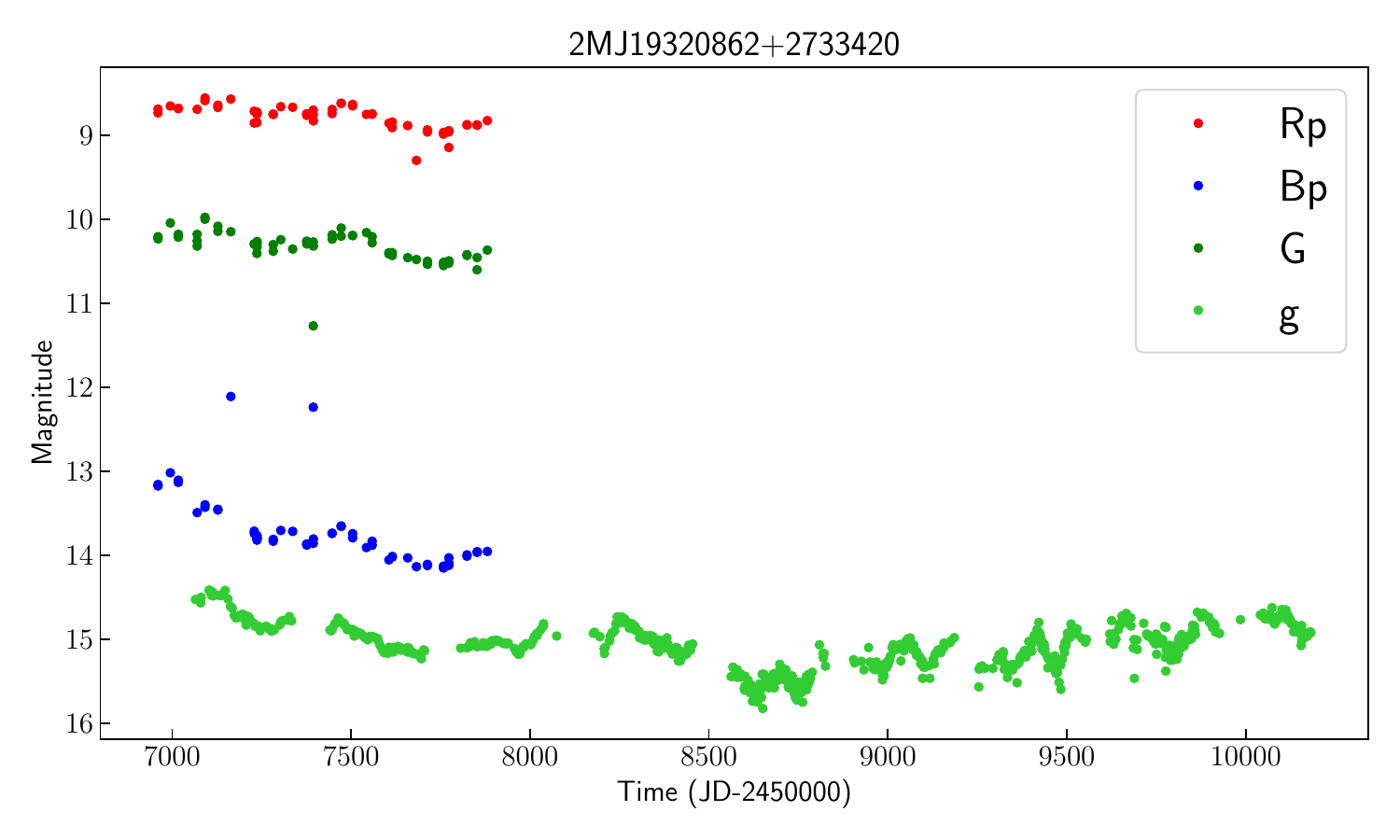}\hspace{-0.5em}
    \includegraphics[width=0.5\textwidth]{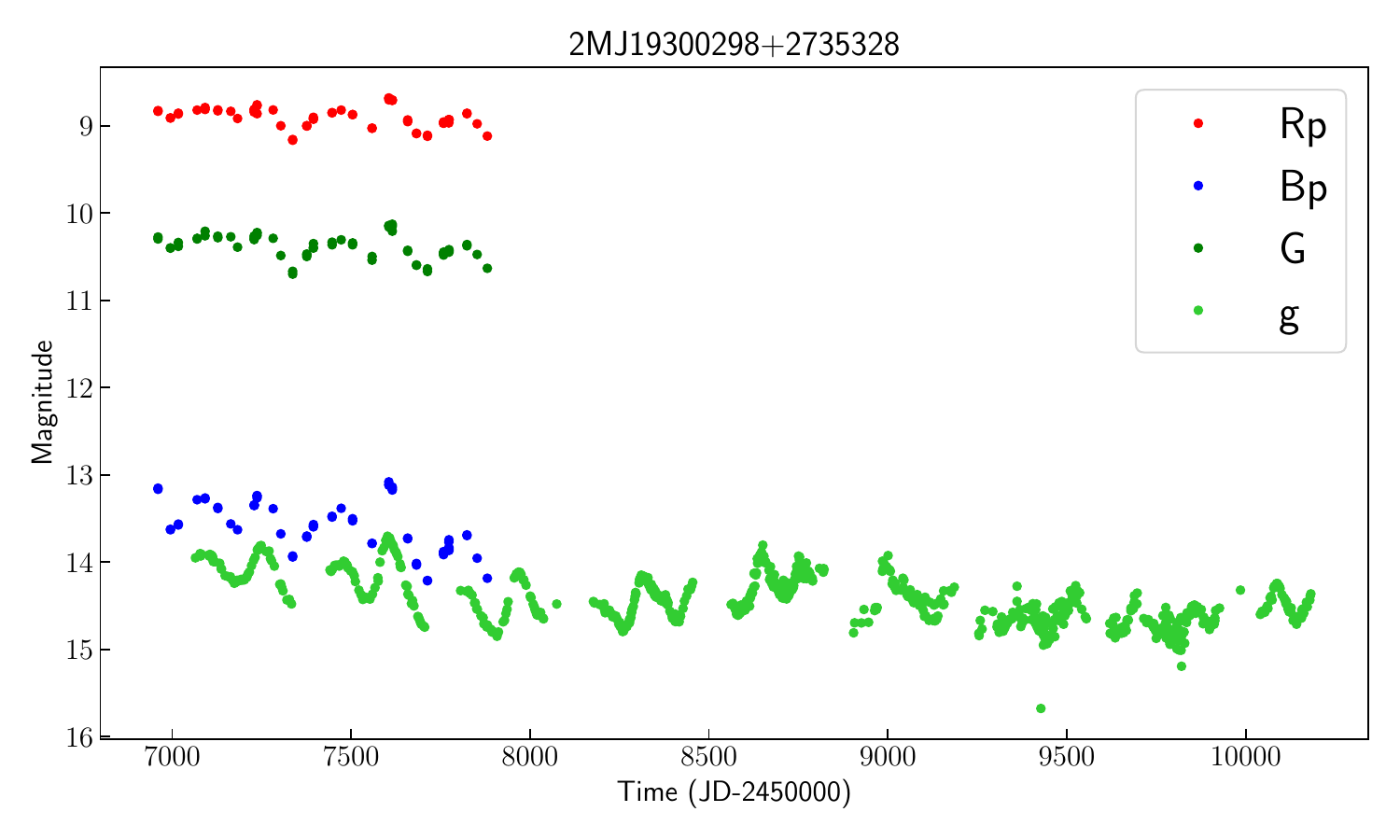}
    \caption{ASAS-SN and Gaia DR3 light curves, labeled by their 2MASS ID. Top left: The Symbiotic variable V2873 Cyg with both $|\Delta M_{G}|$ and $|\Delta(B_{p} - R_{p})| > 0.5$ mag. Top right: An unclassified star in the AGB group with $|\Delta M_{G}|> 0.5$ mag. Bottom: Semi-regular variables NSVS 8366631 (left) and MT Vul (right) with $|\Delta(B_{p} - R_{p})| > 0.5$ mag.}
    \label{fig:gaialcs}
\end{figure}

We used the Gaia light curves to further examine the changes in optical color beyond ASAS-SN. Fig.~\ref{fig:gaia} shows the change in absolute magnitude $\Delta M_{G}$ and color $\Delta(B_{p} - R_{p})$ between the first and last Gaia epochs. Fig.~\ref{fig:gaia} also shows the directions corresponding to changes in temperature, extinction, and luminosity. We see that the typical changes are roughly parallel to changes in temperature. There are 34 with changes of $|\Delta M_{G}|$ or $|\Delta(B_{p} - R_{p})| > 0.5$ mag (after removing any visually spurious points). Almost all of these candidates belong to the AGB group and their shifts across the CMD are shown in Fig.~\ref{fig:gaia}. Fig.~\ref{fig:gaialcs} shows 2 examples with $|\Delta M_{G}| > 0.5$ mag and 2 with $|\Delta(B_{p} - R_{p})| > 0.5$ mag.

\begin{figure}[!tb]
    \centering
    \includegraphics[width=\textwidth]{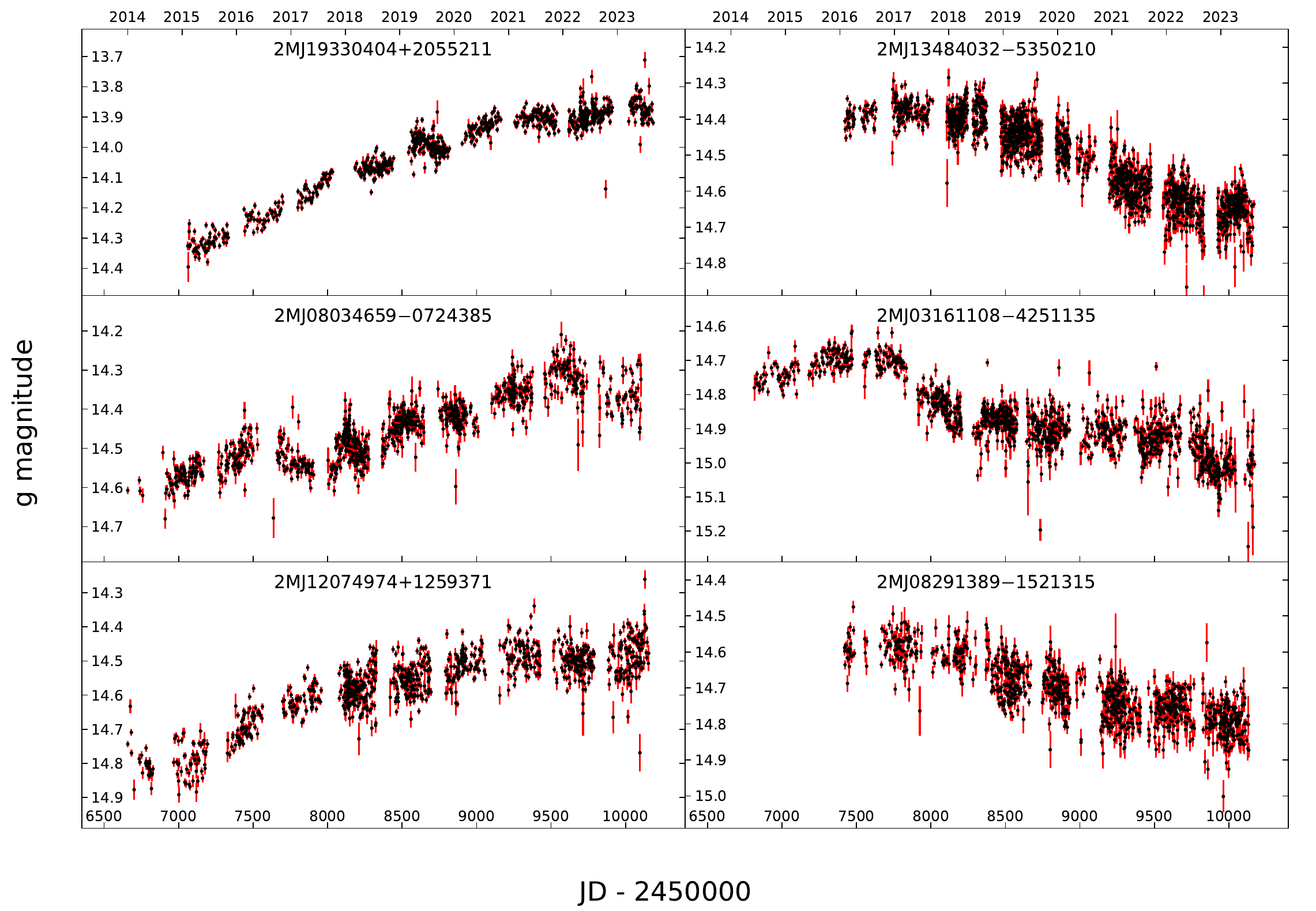}\\
    \caption{Example light curves for the main sequence group labeled by their 2MASS ID. The left (right) panels show the 3 brightening (dimming) sources with the largest (top), median (middle), and smallest (bottom) slopes in the sample. The time axes are the same, but the magnitude ranges vary by object.}
    \label{fig:ms}
\end{figure}

There are 61 candidates in the main sequence group, including 20 showing regular periodicity, growing the total number of main sequence slow variables to 240.  
As seen in PK25, their mean brightness changes are fairly linear, with modest bumps and wiggles. They are concentrated towards the lower main sequence in the CMD (Fig.~\ref{fig:cmdgroups}),
with periods generally greater than $\sim 3$~days when periodic. They largely lie below the Kraft break \citep{Kraft1967}, which suggests that rotation,
convection and magnetic activity are driving the variability.  
The behavior of this group does not seem to be associated with 
dust, since only a few have red mid-IR colors or significantly
opposite signs for the slopes of the optical flux and the mid-IR color.
The Gaia evolution of this group is mostly consistent with changes in temperature, with only one candidate having a $|\Delta M_{G}| > 0.5$ mag.
Fig.~\ref{fig:ms} shows six examples of light curves, where we select the three brightening sources and three fading
sources with the minimum, median, and maximum amplitudes based on the larger of the slope and $\Delta g/10$~years. Additionally, when possible, we choose sources that have light curves that span at least 9 years to display examples with the longest baselines. The typical example has a relatively steady trend extending over the decade of observation, but almost all show evidence of an extremum, which typically suggests that this is also roughly the timescale for starting to revert to the mean. 
The typical amplitudes of variability are smaller than found in the other groups, but a few have large magnitude changes, such as the example in the top left panel. This unclassified star maintains a roughly constant change in magnitude over the course of 9 years with a total magnitude change of $\sim 0.6$ mag.

\begin{figure}[!tb]
    \centering
    \includegraphics[width=\textwidth]{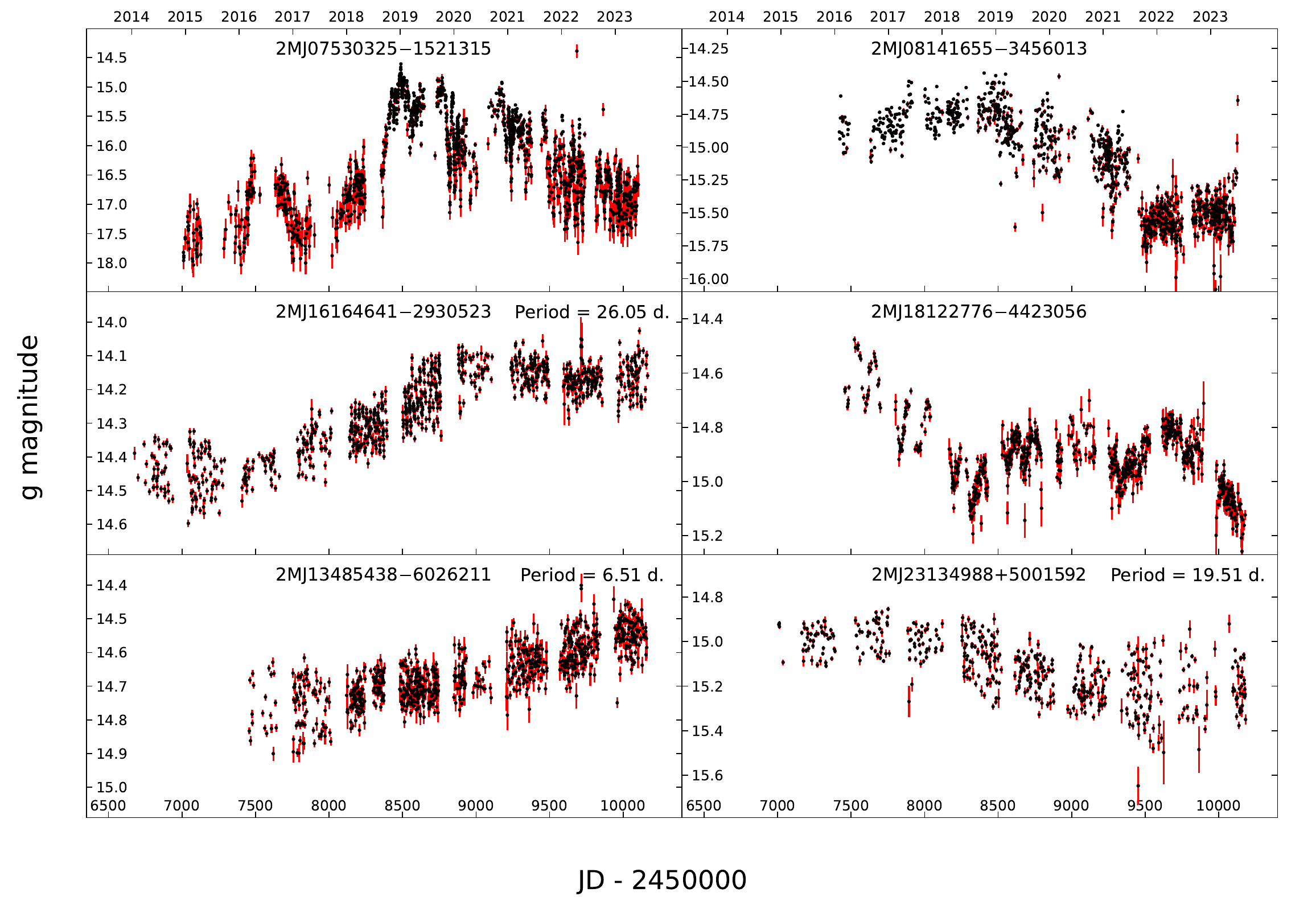}\\
    \caption{Example light curves for the subgiant/giant group. The format is the same is in Fig.~\ref{fig:ms}. If a period exists, it is given in the upper right corner.}
    \label{fig:g}
\end{figure}

The subgiant/giant group consists of 120 stars and exhibit behavior similar to that of the main sequence group, typically having monotonically increasing and decreasing light curves. Adding these to the candidates found in PK25, this group grows to 454 total candidates. These stars tend to exhibit larger variability amplitudes in each season compared to their main sequence counterparts and a have a larger fraction of candidates with periodicity (348 of 454). 
In the Gaia CMDs, they lie towards the red edge of the $B_P-R_P$ color distribution both relative to all stars and the giant branch (Fig.~\ref{fig:cmdgroups}).
They roughly lie in the region occupied by RS~CVn and (sub)sub-giant rotational variables (see \citealt{Anya2024}).
88 of these candidates are flagged for dust related variability.  
Only one new candidate is classified in \texttt{SIMBAD} (see Table~\ref{tab:cands}) as a YSO.
Like the main sequence group, the Gaia color evolution of this group is mostly consistent with changes in temperature, with only 3 of the candidates having a $|\Delta(B_{p} - R_{p})|$ or $|\Delta M_{G}| > 0.5$ mag.
Fig.~\ref{fig:g} shows 6 example light
curves.
The typical example for this group generally has a steady trend over the decade of observation with shorter periodic variability. Like the main sequence group, many show evidence of an extremum. The top panels display examples of stars with more extreme variability. 
The example in the top left has one of the largest brightness changes in the whole sample, brightening by $\sim 3.5$ magnitudes over 4 years before fading by $\sim 2.5$ mag over the next 5 years. The example in the top right brightens and fades slowly over 5 years before abruptly fading by $\sim 0.75$ mag.

\begin{figure}[!tb]
    \centering
    \includegraphics[width=\textwidth]{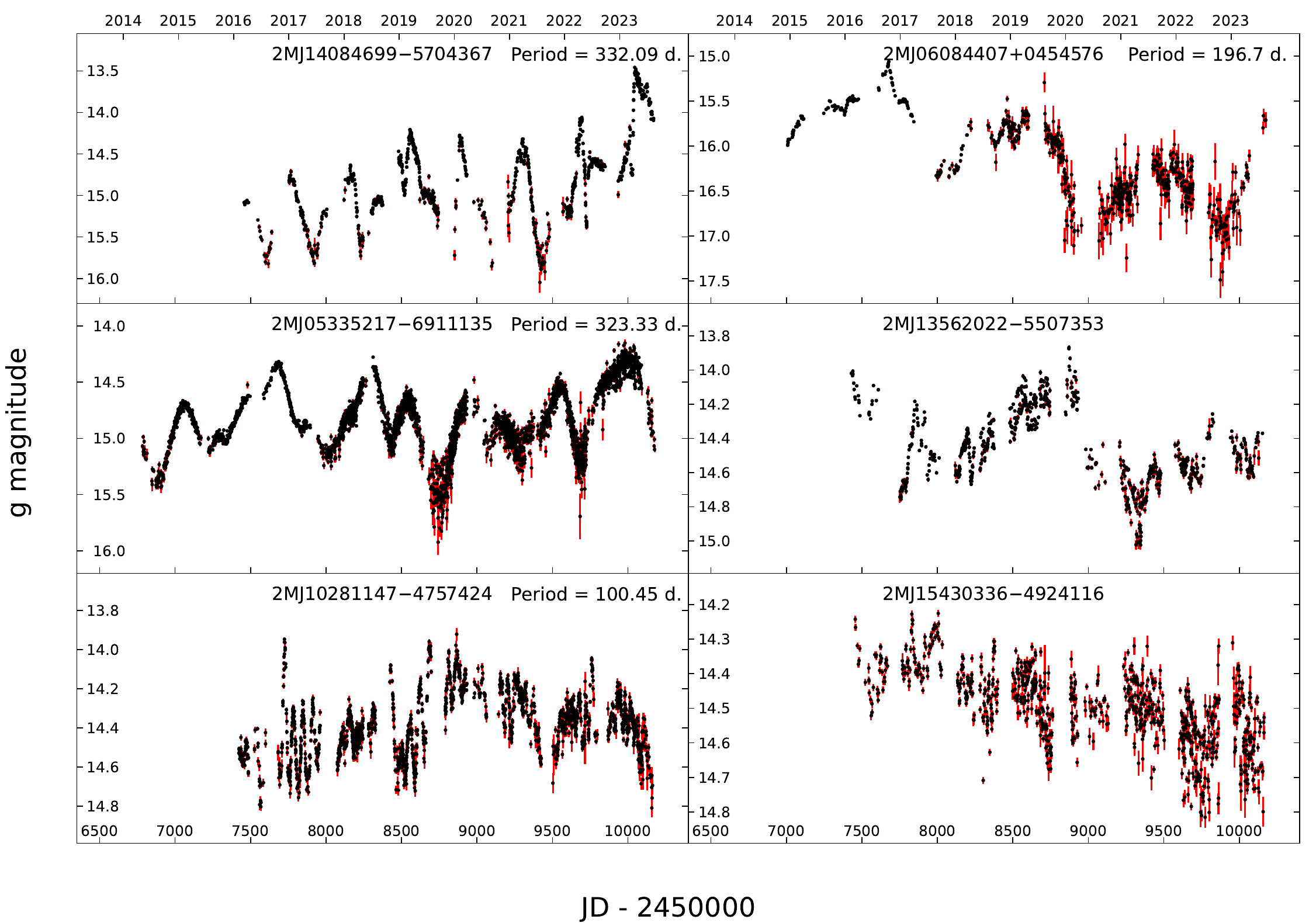}
    \caption{Example light curves for the AGB group. The format is the same is in Fig.~\ref{fig:ms}.}
    \label{fig:agb}
\end{figure}

The AGB group is made up of 205 candidates with $14.5<g<15$ mag, leading to a total of 408 candidates in the full sample including PK25. This group shows distinct differences from the first two groups, having larger amplitudes of variability within  each season, and nearly all having evidence of periodicity (346 of 408).
This group contains all 21 stars flagged in {\tt SIMBAD} as Carbon stars with $14.5<g<15$ mag (see Table~\ref{tab:cands}).
As noted earlier, they tend to be modestly
more luminous than both randomly selected variable stars in that part of the CMD (Fig.~\ref{fig:twins}).
Only 6 from this group have a large mid-IR excess or appear to have dust-related variability.
The Gaia evolution of the AGB group is more scattered, with most of the candidates lying outside of the tight correlation seen for the other groups. This group has the most candidates with evident brightness variations in Gaia, with 22 having $|\Delta(B_{p} - R_{p})|$ or $|\Delta M_{G}| > 0.5$ mag.
Fig.~\ref{fig:agb} shows 6 example light
curves, selected in the same way as for Fig.~\ref{fig:ms}.
The typical examples of this group have continuous irregular variability and slow brightness trends and much longer periods than the other groups. 
The examples with the largest overall long term variability are in the top panels. 
The top left panel has a SR variable (V0710 Cen) with a period of $\sim$ 332 days that brightens $\sim 2$ mag over $\sim 8$ years. The top right panel displays a MISC variable (ASAS J060844+0455.0) with a period of $\sim$ 197 days that has multiple $\sim 1$ mag dimming events with a total change in median brightness of $\sim 2.5$ mag over 10 years.

\begin{figure}[!tb]
    \centering
    \includegraphics[width=\textwidth]{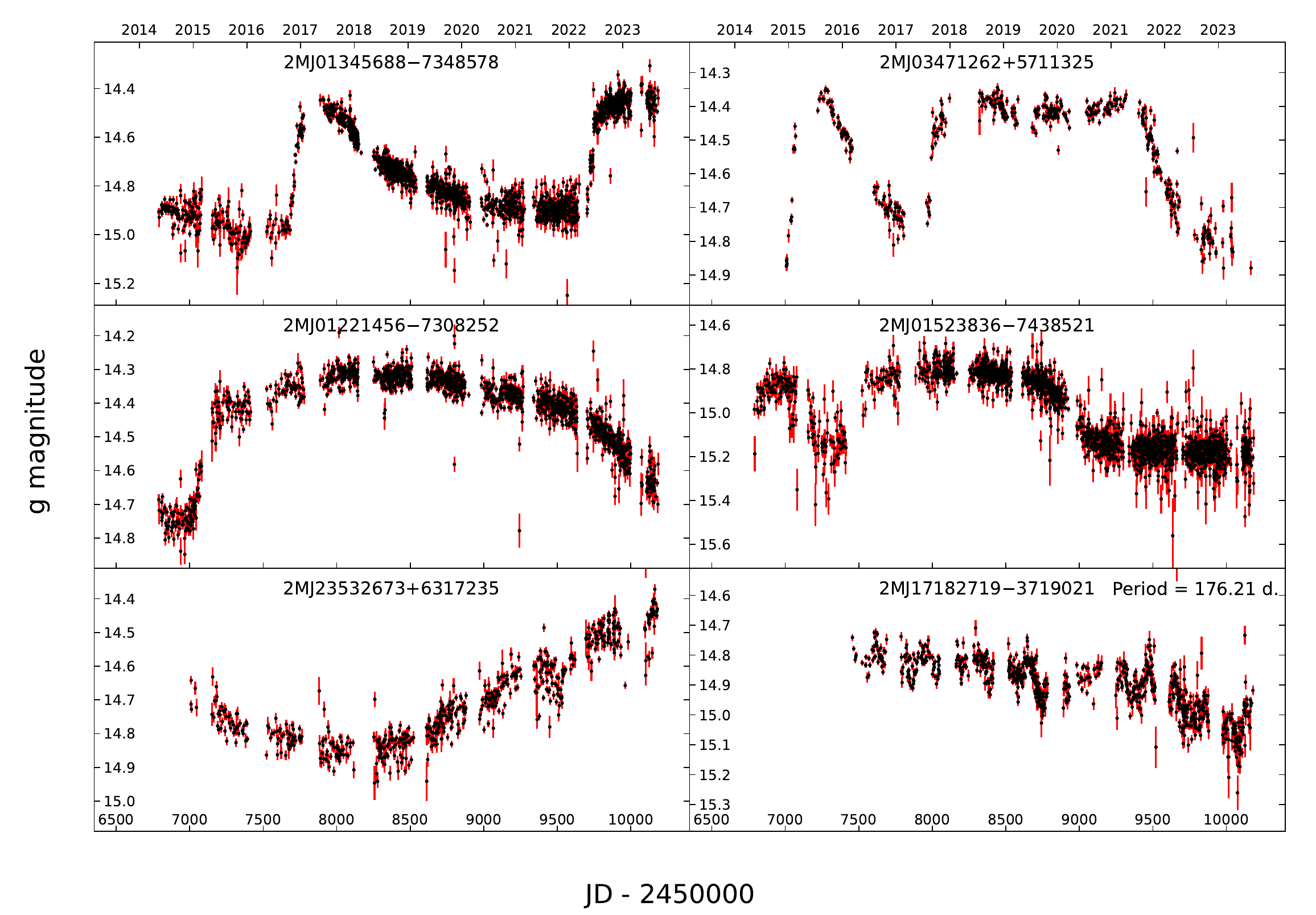}\\
    \caption{Example light curves for the blue luminous group. The format is the same is in Fig.~\ref{fig:ms}.}
    \label{fig:blue}
\end{figure}

With 20 new candidates with $14.5<g<15$ mag, the luminous blue group is still one of the smallest groups of the sample, containing only 63 total candidates.
This group mainly contains stars with sudden jumps or dips in their light curves, which are very different from the trends displayed in other groups. 
Most of the group displays no periodicity, with only one new periodic candidate. 
Many of the candidates appear to have eruptive variability similar to the behavior of Be stars (e.g., \citealt{Labadie2017,Figueiredo2025}). 
We cross matched the new stars with the Be Star Spectra Database (BeSS, \citealt{BeSS}) and found only 2 matches, but 7 are Be (2) or emission line (5) stars in \texttt{SIMBAD} (see Table~\ref{tab:cands}).
There is no obvious class of existing variables to use for comparisons of amplitudes and periods. Almost half of this group (27 out of 63)  were flagged as having dust driven variability, such as the examples in Fig.~\ref{fig:wise2}.
The Gaia magnitude evolution of most of this group is relatively small. One candidate had a $|\Delta M_{G}| > 0.5$ mag. Candidates in this group deviate more from the tight line correlation between $\Delta(B_{p} - R_{p})$ and $\Delta M_{G}$ set by the main sequence and subgiant/giant groups (Fig.~\ref{fig:gaia}), containing a few sources that become more luminous over the duration of their light curves.
Fig.~\ref{fig:blue} shows 6 example light
curves, selected in the same way as for Fig.~\ref{fig:ms}.
Typical examples for this group contain more abrupt brightening or dimming events and their amplitudes of variability are occasionally extreme, but not as large as the amplitudes of the AGB group. 
The most extreme examples of variability in this group are shown in the top panels.
The $\gamma$ Cassiopeiae eruptive irregular variable (ASASSN-V J-13456.88-734857.9) in the top left panel undergoes two large short brightening events of $\sim 0.6$ mag before dimming again over 10 years. 
The rapid irregular variable (Mis V1325) in the top right panel has several brightening and dimming events of $\sim 0.5$ mag over the total duration of its light curve.

\begin{figure}[!tb]
    \centering
    \includegraphics[width=\textwidth]{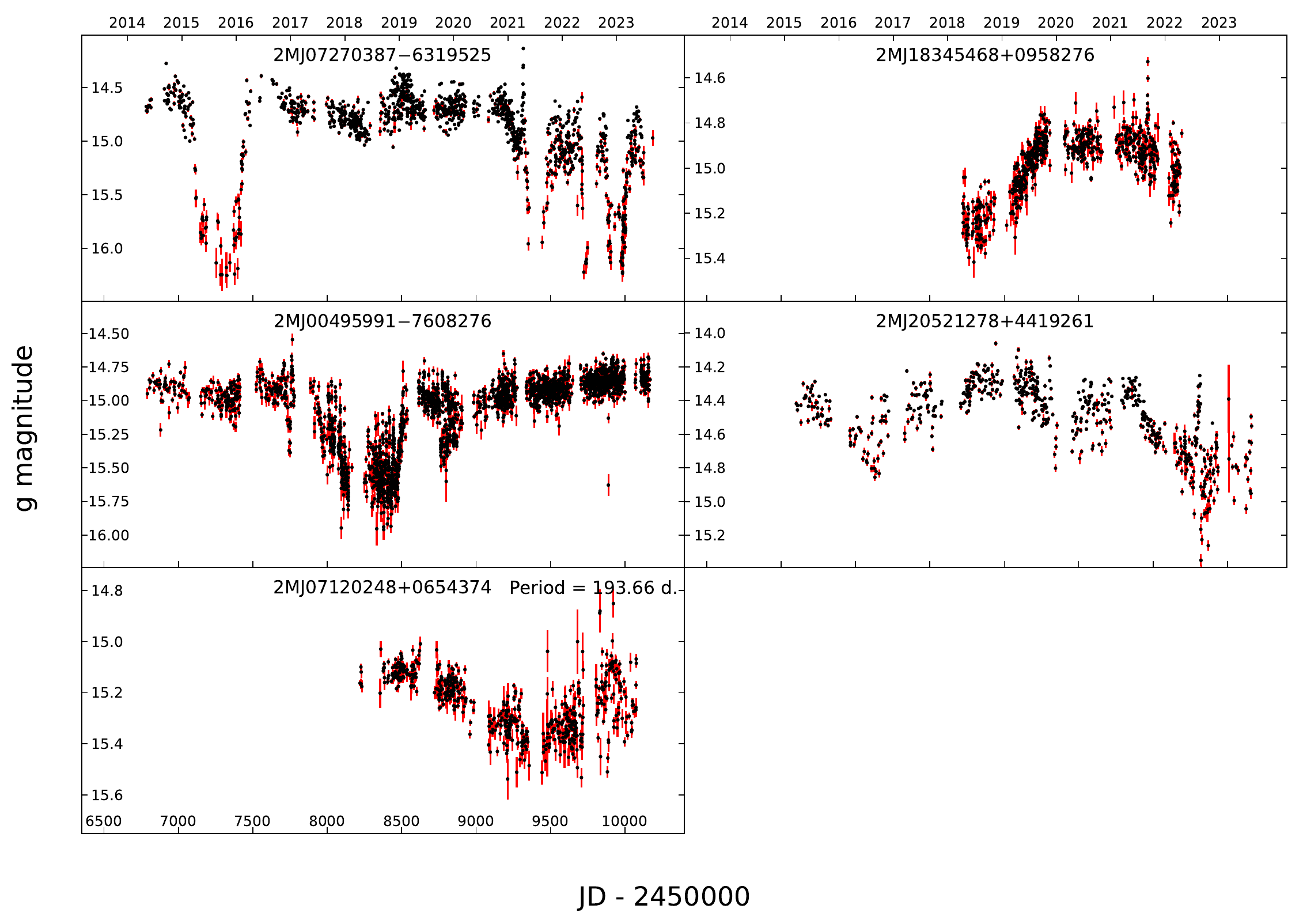}\\
    \caption{Example light curves for the novae group. The format is the same is in Fig.~\ref{fig:ms}.}
    \label{fig:cat}
\end{figure}

The novae group sits to the left of the main sequence and
contains only 5 new candidates, leading to a total of 11 candidates over the full sample. The new candidates in this group are made up almost entirely of previously classified cataclysmic variables. 
Two of these are ``anti-dwarf'' novae (NL/VY variables
in AAVSO) showing short, deep dimming episodes.
The fadings in the anti-dwarf novae are thought to be caused by low rates of mass transfer from their luminous white dwarf companions \citep{Warner1995,Kato2002}. These candidates appear in the top and middle left panels.
Another 2 of these candidates are previously classified as ``Nova-like'' variables (NL or UX in AAVSO) showing ``flickering''-small short brightening and dimming events. These variables are generally all non-eruptive CVs \citep{Warner1995,Bruch2000}.
These candidates appear in the top and middle right panels.
The last candidate in this group is not classified in any catalogs, but shows periodicity.
Two of the new candidates were flagged as having dust related variability.
Over the full sample, none of these candidates show significant Gaia color evolution.
Fig.~\ref{fig:cat} shows all 5 of the new candidates.

\begin{figure}[!tb]
    \centering
    \includegraphics[width=\textwidth]{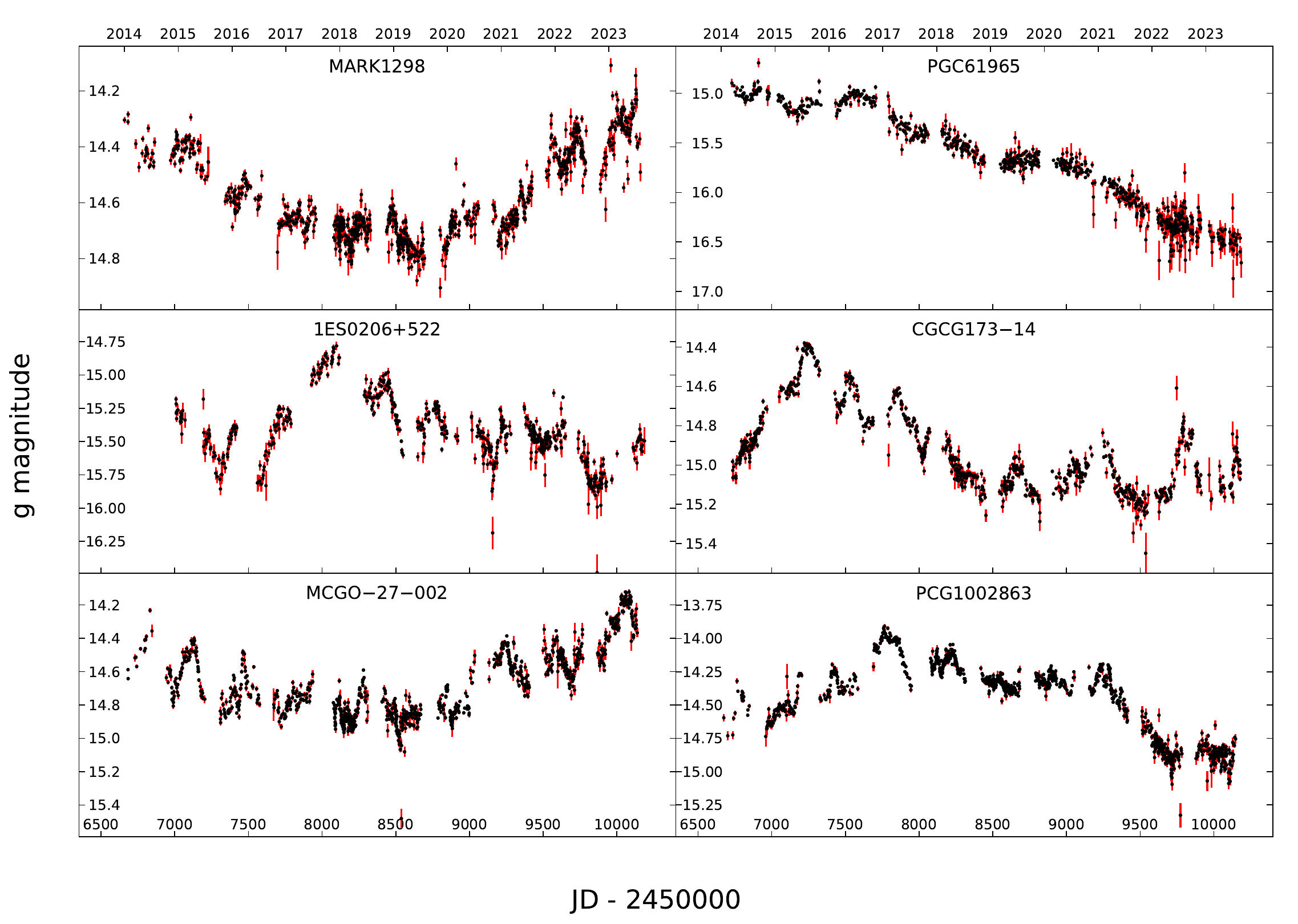}\\
    \caption{Light curves for all 6 AGN.}
    \label{fig:agn}
\end{figure}

Six of our new candidates are known AGN. 
These AGN have variability within each observing season and throughout the total duration of their light curves.
Fig.~\ref{fig:agn} shows their light
curves. The most dramatic brightness changes are seen in the top right panel, where the z=0.12 AGN PGC61965 fades by $\sim 2$~mag over 10 years. The remaining five
(2MJ1129$-$0424/MARK1298, 2MJ0209$-$5226/1ES0206$+$522, 2MJ1835$-$3241/CGCG173$-$14, 2MJ1021$-$0327/MCGO$-$27$-$002, and 2MJ1109$-$0830/PGC1002863) all have $z<0.1$.

\begin{deluxetable*}{llcccccccc}
\tablecaption{Long-term variability candidates and their properties. The full table is available in the electronic version of the paper. The classifications and spectral types are from \texttt{SIMBAD}.\label{tab:cands}}
\tablewidth{0pt}
\tablehead{
\colhead{2MASS} & \colhead{$g$} & \colhead{Optical Slope} & \colhead{$\Delta g$} & \colhead{Group} & \colhead{Period} & \colhead{W1$-$W2} & \colhead{W1$-$W2 Slope} & \colhead{Class} & \colhead{Spec. Type} \\
\colhead{} & \colhead{(mag)} & \colhead{(mag/yr)} & \colhead{(mag)} & \colhead{} & \colhead{(days)} & \colhead{(mag)} & \colhead{(mag/yr)} & \colhead{} & \colhead{}
}
\startdata
2MJ06400303$+$1800009 & 14.506 & $-0.047$ & 0.374 & Subgiant/giant & \ldots & $-0.031$ & $-0.003$ & \ldots & \ldots \\
2MJ08453329$-$3923085 & 14.505 & $+0.060$ & 0.422 & AGB & 204.686 & $+0.638$ & $+0.020$ & \ldots & M \\
2MJ07592735$-$3556579 & 14.500 & $+0.038$ & 0.579 & AGB & 363.140 & $+0.058$ & $-0.064$ & \ldots & S?p \\
2MJ20133295$+$1515111 & 14.505 & $+0.079$ & 0.489 & AGB & 275.192 & $-0.056$ & $+0.002$ & \ldots & \ldots \\
2MJ05312358$+$1209438 & 14.501 & $-0.050$ & 0.397 & Main Sequence & 222.475 & $+0.629$ & $+0.003$ & \ldots & K0Ve \\
2MJ13364117$-$5415136 & 14.500 & $+0.033$ & 0.230 & Main Sequence & \ldots & $-0.036$ & $-0.004$ & EB & \ldots \\
2MJ15250533$-$6429208 & 14.505 & $-0.033$ & 0.210 & AGB & \ldots & $-0.355$ & $-0.014$ & \ldots & \ldots \\
2MJ13484032$-$5350210 & 14.506 & $+0.044$ & 0.308 & Main Sequence & \ldots & $-0.052$ & $-0.003$ & \ldots & \ldots \\
2MJ13572359$-$5500237 & 14.504 & $-0.012$ & 0.432 & AGB & 98.990 & $-0.446$ & $-0.055$ & \ldots & M7 \\
2MJ22283374$+$5402576 & 14.502 & $-0.034$ & 0.309 & Blue Luminous & \ldots & $+0.087$ & $+0.023$ & \ldots & \ldots \\
\enddata
\end{deluxetable*}

\needspace{5\baselineskip}
\section{Conclusion}
\label{sec:concl}

We find 426 new slow variable candidates among the 5,685,060 ASAS-SN sources with $14.5<g<15$ mag. They were chosen to have mean variability rates of $\gtorder 0.03$ mag/year over $\sim 10$ years. This sample includes 226 previously classified variables and 200 variables identified for the first time. 
The existing variability classifications in the literature do not capture the slow variability behavior of these candidates.
Following PK25, we group these candidates into 5 groups, finding that most are AGB stars rather than subgiants and giants.
We find that 259 candidates have periodicity, most having periods longer than 10 days and belonging to the AGB group.
With WISE light curves, we find that 49 candidates with mid-IR brightness changes opposite to the optical, suggestive of changes in circumstellar dust and 17 candidates with a large mid-IR excess. Based on their shifts in a Gaia CMD, the variability is largely driven by temperature changes. One candidate in our overall sample is also found in a systematic search for ASAS-SN stars showing light curve dips due to being a very long period eclipsing binary or optically thick occulting dust \citep{Brayden25}.
The full list of variables is given in Table~\ref{tab:cands} and their light curves can be obtained using ASAS-SN SkyPatrol 2.0 (http://asas-sn.ifa.hawaii.edu/skypatrol/).

The main sequence candidates lie below the \cite{Kraft1967} break, have the smallest slope amplitudes of the entire sample, and generally have the shortest periods if periodic.
Eleven are flagged as possibly having dust related variability. Much like their brighter counterparts, their activity likely represents the extremes of star spot changes on longer time scales.
Only one new candidate is flagged as a YSO in \texttt{SIMBAD}. 
The subgiants/giants fall on the redder edge of the color distribution of RGB stars 
and contain candidates with mainly intermediate length periods. 
Stars in this group are likely associated with RS CVn/(sub)sub-giant rotational variables. However, 23 percent of these candidates are flagged as having dust-related variability.
The AGB group stars seem to be slightly more luminous than randomly selected L and SR variables. Many of these candidates have high amplitude, slow ($\geq$ 50 days) periodic variability as well. 
Only 2 show a significant large mid-IR excess, implying that their variability is not driven by dust.

The variability of the blue stars is more eruptive, with high-amplitude variability and almost no periodicity.
Their behavior is similar to the behavior of Be stars, however only 9 are flagged as Be or emission line stars in BeSS \citep{BeSS} and \texttt{SIMBAD}. A large fraction of this group is flagged as having dust related variability.
The novae group has deep fadings spanning several magnitudes and is almost entirely made up of different cataclysmic variables.  
Finally, we find a small number of AGN (6), where the highest amplitude source is the z=0.12 AGN PGC61965 which has steady fading by $\sim 1.5$ mag over the last 10 years.

Comparing these fainter candidates to the 782 slow variable candidates found with 13$<$g$<$14.5 mag in PK25, we find that sources categorized into each group display similar behaviors and trends based on their CMD position and infrared variability. Though our first set of candidates in the overall sample mainly sit in the subgiant/giant group, the candidates in this work are mostly a part of the AGB group, increasing the overall number of longer period pulsating variables. With 426 added variables from this work between 14.5$<$g$<$15 mag we can extend our total number of slow variable candidates to 1208. 

As these variables appear to be relatively rare, to continue our exploration of long term variability, we plan to extend our list of candidates over the magnitude parameter space by continuing the search for slow variability in both brighter saturated stars (see \citealt{Winecki2024}) and stars dimmer than 15 mag. As ASAS-SN's baseline continues to grow, we plan to expand our search to explore smaller amplitudes of variability, develop methods to significantly improve false positive rejection as well as continue to explore and characterize the growing population of deep eclipses in ASAS-SN (\citealt{Joey2016, Raquel25}).

\section*{Acknowledgments}

CSK and KZS are funded by NSF grants AST-2307385 and AST-2407206.
ASAS-SN is funded by Gordon and Betty Moore Foundation grants GBMF5490 and GBMF10501 and the Alfred P. Sloan Foundation grant G-2021-14192.
The Shappee group at the University of Hawai‘i is supported with funds from NSF (grants AST-2407205) and NASA (grants HST-GO-17087, 80NSSC24K0521, 80NSSC24K0490, 80NSSC23K1431).
S.D. is supported by the National Natural Science Foundation of China
(Grant No. 12133005).  S.D. acknowledges the New Cornerstone Science
Foundation through the XPLORER PRIZE.
We acknowledge ESA Gaia, DPAC and the Photometric Science Alerts Team (http://gsaweb.ast.cam.ac.uk/alerts).
This research has made use of the SIMBAD database,
operated at CDS, Strasbourg, France. 
We acknowledge with thanks the variable star observations from the AAVSO International Database contributed by observers worldwide and used in this research.
This work has made use of the BeSS database, operated at LESIA, Observatoire de Meudon, France: http://basebe.obspm.fr.

\bibliographystyle{mnras}

\bibliography{oja_template}




\end{document}